\documentclass[preprint,aps,prc,tightenlines,floatfix,showpacs]{revtex4}
\usepackage{graphicx}
\usepackage{bm}
\usepackage{amsmath}
\usepackage{amssymb}
\begin{document}
\preprint{}
\title{Probing Correlated Ground States with Microscopic Optical Model for 
Nucleon Scattering off Doubly-Closed-Shell Nuclei.}
\author{M. Dupuis}
\author{S. Karataglidis\footnote[1]{Present adress: School of Physics, 
University of Melbourne, Victoria 3010, Australia.}} 
\author{E. Bauge}
\author{J.P. Delaroche}
\author{D. Gogny\footnote[2]{Present adress: Lawrence Livermore National
Laboratory, California, USA.}}
\affiliation{Commissariat \`a l'Energie Atomique, D\'epartement de
  Physique Th\'eorique et Appliqu\'ee, Service de Physique
  Nucl\'eaire, Boite Postale  12, F-91680 Bruy\`eres-le-Ch\^atel, France}

\date{\today}
\begin{abstract}
The RPA long range correlations are known to play a significant role in 
understanding the depletion of single particle-hole states observed in (e, e')
 and (e, e'p) measurements. Here the Random Phase Approximation (RPA) theory, 
implemented using the D1S force is considered for the specific purpose of 
building correlated ground states and related one-body density 
matrix elements. These may be implemented and tested in a fully microscopic 
optical model for NA scattering off doubly-closed-shell nuclei. A method is 
presented to correct for the correlations overcounting inherent to the RPA 
formalism. One-body density matrix elements in the uncorrelated 
(i.e. Hartree-Fock) and correlated (i.e. RPA) ground states are then 
challenged in proton scattering studies based on the Melbourne microscopic 
optical model to highlight the role played by the RPA correlations. Effects 
of such correlations which deplete the nuclear matter at small 
radial distance (r $<$ 2 fm) and enhance its surface region, are getting 
more and more sizeable as the incident energy increases. Illustrations 
are given for proton scattering observables measured up to 201 MeV for 
the $^{16}$O, $^{40}$Ca, $^{48}$Ca and $^{208}$Pb target nuclei. Handling
 the RPA correlations systematically improves the agreement between 
 scattering predictions and data for energies higher than 150 MeV.
\end{abstract}
\maketitle


\section{Introduction}
Our understanding of the many facets of the nuclear structure properties has 
been and still is relying on the picture of independent particles moving in a 
mean potential. This picture stands at the foundation of the shell model which 
nowadays serves routinely as the basis of nuclear structure calculations, and 
is implicit to the self-consistent mean-field (i.e. Hartree-Fock) 
description of nuclear ground states. For independent particle motion, the 
occupancy associated to nucleon orbitals is 1 or 0 depending upon whether the 
single-particle level is below or above the Fermi energy, respectively. 

It is only recently that the quenching of shell model occupation probabilities
 has been disclosed in a dedicated series of experiments in which incident 
electrons serve to map detailed structure properties hard to reach using other
 probes. First hints revealing such a quenching  came in measurements of 
electrons scattering from $^{206}$Pb and $^{205}$Tl, from which the 
$3s_{1 \! / 2}$ proton radial wave function was determined.
Its shape is peaked in the central region, and close to expectations for a 
$3s_{1 \! / 2}$ wave function. Minor adjustment of the $3s$-hole strength 
provided an improved data prediction \cite{Cave}. Evidence for partial 
occupancy for this orbital was provided later on from a joint analysis of
(e, e') and (e, e'p) experiments. The $3s_{1 \! / 2}$ orbital was found to 
be depleted by a $(18 \pm 9) \%$ amount \cite{Qui1,Qui2}. Today the absolute 
occupation probability of this proton orbital is evaluated to be 
$0.76 \pm 0.07$ \cite{Sick}.

Further detailed information on the single-particle structure have been 
recently gained through measurements of the spectral function $S(E,k)$, 
where $E$ and $k$ are the removal energy and momentum, respectively, of 
a proton in (e, e'p) knockout experiments. For $^{208}$Pb, these measurements 
performed at high binding energy and momentum transfer show that mean-field 
predictions are lying far below the data, highlighting the need for 
consideration of tensor \cite{Pan3,Fant2} as well as short- \cite{Lapi,Piep2,Heis} and 
long-range correlations beyond the mean field \cite{Bob1,Bob2,Pan1,Pan2}. 
A wealth of methods and models have been adopted to tackle this issue. These 
are the Green's functions method \cite{Wiri,Piep}, the variational Monte-Carlo method 
\cite{Wiri2,Pudl}, 
the correlated basis function theory \cite{Fant}, the particle-vibration 
model \cite{Bern}, the dispersive optical model extrapolated to bound state 
region \cite{Mah1,Mah2,Mah3,Mah4}, and the Random Phase Approximation 
\cite{Dec1,Dec2,Lens,Gog1,Ma}. Among the correlations which have been 
considered so far, the long-range ones appear important for curing the 
deficiencies associated with the mean-field predictions.

In the present work we investigate the impact that nuclear long-range 
correlations have on the interaction of nucleons incident on 
doubly-closed-shell nuclei, among which $^{208}$Pb, a nucleus for which 
many scattering observables have been measured. In the past, detailed 
experimental information on nuclear structure as gained from electron 
scattering measurements played a key role in building effective 
NN forces and mass operators for nucleon scattering studies in the folding 
model framework \cite{Petro,procLove}. Now, that a successful NA microscopic optical model 
(OM) based on a g-matrix interaction has been established in r-space 
\cite{Am00}, it is timely to push the limits of its predictive power 
using various microscopic picture information. Several studies along 
this line have already been published. For example, no core shell model 
wave functions have been adopted in successful interpretations of proton 
scattering measurements for $^{12}$C and light nuclei below and at the 
neutron drip-line \cite{Geramb,Na98,Schw2}. Hartree-Fock predictions based on Skyrme 
forces have also been challenged in proton and neutron elastic scattering 
studies at medium energy to provide estimates of neutron skin thickness in 
$^{208}$Pb. Here, the correlated ground states of stable doubly-closed-shell nuclei, 
built using the finite range, density dependent D1S force \cite{BGG} in 
the self-consistent RPA theory \cite{BG1}, are used instead and thoroughly 
tested.

Our paper is organized as follow. The main features of the fully antisymmetric,
 microscopic NA optical model are described in Sec.\ref{melbourne}. Section 
\ref{structure} includes a brief presentation of the HF+RPA theory for 
establishing our notations, and describes the method used to fix the 
well-known double counting problem. RPA predictions are compared to 
experimental data for charge and neutron radial shapes of $^{208}$Pb in its 
ground state. One body density matrix elements in the correlated ground state 
are then provided. Finally, optical model predictions based on HF and HF+RPA 
one body density matrix elements are compared in Sec.\ref{analyse} to various 
scattering observables in the 40-201 MeV incident proton energy range for 
$^{16}$O, $^{40}$Ca, $^{48}$Ca and $^{208}$Pb, and to scattering predictions 
based on the Skyrme SkM$^*$ \cite{guet} force.

  \section{Microscopic optical potential from the Melbourne $g$ matrix}
  \label{melbourne}

  The full details  of the Melbourne $g$ matrix  optical potential may
  be  found in  Ref.~\cite{Am00}, to  which  we refer  the reader.  We
  present  a  brief  summary  of  the  derivation  of  the  potential,
  highlighting those  points relevant to  the use of RPA  densities in
  its calculation and the observables obtained therefrom.

  In  folding models  of  the  optical potential,  one  starts with  a
  credible effective  $NN$ interaction. In  the case of  the Melbourne
  potential, the effective $NN$  interaction is the $g$ matrix derived
  from  the Bonn-B  $NN$ interaction  \cite{Ma87}. The  $g$ matrix for
  infinite  matter  is  a  solution  of  the  Bruckner-Bethe-Goldstone
  equation in momentum space, \textit{viz.}
  \begin{equation}
    g\left( \mathbf{q}', \mathbf{q}; \mathbf{K} \right) = V\left(
    \mathbf{q}', \mathbf{q}\right) + \int V\left( \mathbf{q}',
    \mathbf{k}' \right) \frac{ Q \left( \mathbf{k}', \mathbf{K}; k_f
    \right) }{ \left[ E\left( \mathbf{k}, \mathbf{K} \right) - E
    \left( \mathbf{k}', \mathbf{K} \right) \right] } g\left(
    \mathbf{k}', \mathbf{q}; \mathbf{K} \right) \; d\mathbf{k}'\; ,
  \end{equation}
  where $Q(\mathbf{k}',\mathbf{K};k_f)$ is a Pauli operator and medium
  effects are  included in the  energy denominator. Effective  $g$ 
  matrices are  obtained in coordinate  space for finite  nuclei whose
  Fourier transforms  best map  those momentum space  solutions. Those
  $g$ matrices  so  obtained  contain  central,  tensor,  and  two-body
  spin-orbit terms.  They also are constructed over  all two-body spin
  and isospin  channels, allowing for  a self-consistent specification
  of  proton  and  neutron  scattering,  as well  as  charge  exchange
  reactions. Those $g$ matrices are  then folded with the ground state
  density matrix  elements to give  the optical potential  for elastic
  scattering.

  The optical potential derived therefrom can be cast in the form
  \begin{align}
\label{uopt}
    U( \mathbf{r}, \mathbf{r}'; E ) & = \delta\left( \mathbf{r} -
    \mathbf{r}' \right) \sum_{\alpha\beta} \rho_{\alpha\beta} \int \varphi^{\ast}_{\alpha}(\mathbf{s})
    g_D( \mathbf{r}, \mathbf{s}; E ) \varphi_{\beta}( \mathbf{s} ) \;
    d\mathbf{s} \nonumber \\
    & \; \; \; + \sum_{\alpha\beta} \rho_{\alpha\beta} \varphi_{\alpha}^{\ast}( \mathbf{r} ) g_E(
    \mathbf{r}, \mathbf{r}'; E ) \varphi_{\beta}( \mathbf{r}' ) \nonumber \\
    & = U_D( \mathbf{r}; E ) \delta\left( \mathbf{r} - \mathbf{r}'
    \right) + U_E( \mathbf{r}, \mathbf{r}'; E )\; ,
  \end{align}
  where the subscripts  $D$ and $E$ designate the  direct and exchange
  contributions, respectively. Nuclear structure information enters in
  via the one-body matrix elements $\rho_{\alpha \beta}$ and in  the specification  of the
  bound state single particle wave functions $\varphi_{\alpha}$ and $\varphi_{\beta}$.
  In terms of the RPA (or HF) ground state $\left| 0 \right\rangle$, that
  density matrix element is $\rho_{\alpha \beta} = \langle 0 |
  a^+_\alpha a_\beta |0\rangle$ (see Sec.\ref{structure}).

  The main source of non-locality  in the optical potential is from the
  exchange  term. The  direct  term resembles  a $g\rho$-type  optical
  potential and by definition is  local. The form of the exchange term
  necessarily does not follow this construction: the exchange terms in
  the folding  require that  the sum is  over explicit  effective $NN$
  two-body amplitudes.  As such,  direct comparisons are  not possible
  between  this form  of the  optical  potential and  those which  are
  local, as  constructed from  nonlocal $NN$ amplitudes  through local
  approximations,  or  as  specified  phenomenologically  as  sums  of
  Woods-Saxon form factors.

  To obtain  the observables for scattering, the  optical potential so
  obtained is used  in the nonlocal integro-differential Schr\"odinger
  equation, \textit{viz.}
  \begin{equation}
    \left[ \frac{\hbar^2}{2\mu}\nabla^2 - V_C(r) + E \right]
    \Psi(\mathbf{r}) = \int U(
    \mathbf{r},\mathbf{r}')\Psi(\mathbf{r}') \, d\mathbf{r}'\; ,
  \end{equation}
  where $V_C(r)$  is the Coulomb potential,  and the terms  due to the
  intrinsic spin  of the system  have been suppressed  for simplicity.
  The  code  DWBA98  \cite{Ra98}  is  used to  calculate  the  folding
  potential  from  the effective  $NN$  $g$ matrices  and obtain  the
  relevant scattering observables.
  
  At low  energy, the  averaging over the  coupling to  the nonelastic
  channels represented  by the $g$ matrix  is no longer  valid and the
  derivation  of  the optical  potential  must  be  done in  terms  of
  explicit  channel coupling  to open  and closed  channels.  Such has
  recently  been   constructed  in  terms  of   the  collective  model
  \cite{Am03,Ca04}.

\section{Nuclear Structure}
\label{structure}
As an introduction to this section, it is important to mention that
our approach is not fully consistent. On the one hand, we use the 
g-matrix as an interaction between the projectile and the nucleons in
the target, whereas on the other hand, to calculate nuclear structure,
we consider effective interactions which have been separately
adjusted. As long as we focus on studying medium energy scattering, one
can find justifications for proceeding in this way. However, at low
energy, this approach would be more questionable and it is likely that
the derivation of an optical potential in the frame of a more
fundamental theory (as in \cite{Vill}) should be considered.
 \subsection{The mean field approximation}
 \label{hf} 
The simplest description of the nuclear structure is provided by the 
self consistent mean field theory, which is also called Hartree-Fock (HF). 
There, the ground state is a Slater determinant constructed with individual
particle states which are solutions of the HF equations. In this 
work,  we use the HF results obtained using two different 
interactions. One is the Skyrme SkM* \cite{guet} interaction , and the other 
one is the finite range, density dependent D1S interaction \cite{BGG}. 
The details of the HF formalism used with the D1S density dependent 
interaction can be found in \cite{De80,DGGG}.

In order to calculate the one-body matrix elements $\rho_{\alpha \beta}$ of Sec.\ref{melbourne}, it is convenient to express the HF ground state in 
second quantization as  

\begin{equation}
\label{hfwf}
|HF\rangle =\prod_h a^+_h|0_{HF}\rangle  
\; \; .
\end{equation}  

\noindent The above product contains only occupied states labeled ``h''
(hole states) according to the usual terminology. The creation
operator $a^+_h$ associated with the creation of a hole in a HF single
particle state is defined with: $\varphi_h(r)= \langle r|a^+_h|0_{HF}\rangle $.

 By introducing these notations, the matrix elements $\rho_{\alpha \beta}$ read

\begin{equation}
\rho_{\alpha\beta}=\langle HF|a^+_\beta a_\alpha|HF\rangle 
\; \; .
\end{equation} 
and are diagonal
($\rho_{h,h}=1$ , $\rho_{p,p}=0$) in the HF approximation.

\subsection{Description of the ground state beyond the HF approximation}
\label{rpa}
The density dependent effective interaction D1S has successfully been used 
in various extensions of the mean field theory. Among them, the one of 
interest for our study is the microscopic description of collective 
excitations for closed shell nuclei as described in reference \cite{BG1}. 
We recall some essential features of this approach and make the link with 
the usual RPA theory. This will permit us to define the two variants of 
correlated ground states that we propose for the description of the target.

\subsubsection{Ground state correlations induced by collective excitations}

The approach of \cite{BG1} is based on the quadratic form introduced a 
long time ago to study the stability conditions of the HF solutions. It is 
obtained by performing a Taylor expansion of the energy $E$ up to second 
order in the variation of the density matrix around the equilibrium HF density 
($\rho^{(0)}$). The quadratic form in question is expressed in terms of the 
matrix

\begin{equation}
\label{quadmat}
\left(\begin{array}{cc} A &B\\ B^* & A^*\end{array} \right) 
\; \; , 
\end{equation}
with elements
\begin{equation}
A_{(ph),(p'h')}=\delta_{pp'}\delta_{h,h'}(\epsilon_p-\epsilon_h)
+ \left[ \frac{\partial^2E /\partial\rho_{ph} }
{\partial\rho_{p'h'}}\right]_{\rho=\rho^{(0)}}
\; \; , 
\end{equation}
and
\begin{equation}
\label{Bdef}
B_{(ph),(p'h')}=\left[\frac{\partial^2E}{\partial\rho_{ph}\partial\rho_{h'p'}}
\right]_{\rho=\rho^{(0)}}
\; \; ,
\end{equation}
where $\epsilon_p$ and $\epsilon_h$ are the HF single particle energies for 
a particle state and a hole state, respectively.
This matrix is used to define a set of RPA equations \cite{BG1,Ring}, namely 
\begin{equation}
\label{rpaeq}
\left(\begin{array}{cc} A &B\\ B^* & A^*\end{array} \right) 
\left(\begin{array}{cc} X \\ Y\end{array} \right)
=\omega  \left(\begin{array}{cc} X \\ -Y\end{array} \right)
\; \; ,
\end{equation}
where $\omega$ is a set of eigen-values corresponding to a set of 
eigen-vectors with components $X$ and $Y$. The definition of the matrix 
(\ref{quadmat}) presents the advantage to show that, due to its explicit 
dependence on the density, the particle-hole matrix elements of D1S must 
contain the so-called rearrangement terms besides the usual ones. 
Notice also, that one retrieves the usual particle-hole matrix elements 
when the interaction does not depend on the density. Once such prescription 
is adopted for defining the particle-hole vertices, the approach developed 
in \cite{BG1} follows closely the standard RPA theory as described 
extensively in \cite{Ring}. Below we only give the relevant 
definitions  that introduce the quantities of interest for this work. 
We express the formalism in a representation that accounts for rotational 
invariance and reflection symmetries of the nuclear interaction and the 
mean-field as well (see Appendix \ref{appen1}). Creation and annihilation 
operators are defined through a Bogolyubov transformation

\begin{equation}
\Theta^+_{i,(\pi ,J,M)}=\sum_{p,h} X^{\pi ,J}_{i,(p,h)}A^+_{(p,h)}(\pi ,J,M)
+Y^{\pi ,J}_{i,(p,h)}\bar{A}_{(p,h)}(\pi ,J,M)
\; \; , 
\end{equation}
\[
\bar{\Theta}_{i,(\pi ,J,M)}=\sum_{p,h} 
Y^{\pi ,J}_{i,(p,h)}A^+_{(p,h)}(\pi ,J,M)
+X^{\pi ,J}_{i,(p,h)}\bar{A}_{(p,h)}(\pi ,J,M)
\; \; , 
\]
which mixes the creation and destruction operators, $A^+_{(p,h)}(\pi ,J,M)$ 
and $\bar{A}_{(p,h)}(\pi ,J,M)$ respectively, of independent 
particle-hole pairs with definite angular momentum  and parity. The amplitudes
 $X$ and $Y$ are the components of the solutions of the RPA equations 
defined in (\ref{rpaeq}). Since we work within the quasi-boson approximation, 
the Bogolyubov transformation is nothing but a canonical transformation 
between two sets of bosons. Excitation modes of the nucleus are then defined 
through the action of any creation operator $\Theta^+$ onto the quasi-boson 
vacuum $|\tilde{0}\rangle $ of the destruction operator $\Theta$. This is 
expressed as follows
\begin{equation}
\label{vac1}
|i,(\pi,J,M)\rangle =\Theta^+_{i,(\pi,J,M)}|\tilde{0}\rangle  
\; \; ,
\end{equation} 	
\[ 
\Theta_{i,(\pi,J,M)}|\tilde{0}\rangle =0 \;\;\; \forall\, i,\pi,J,M 
\; \; .
\]
The quasi-boson vacuum can be constructed explicitly from the vacuum 
$|HF\rangle $ of the $A_{(p,h)}(\pi,J,M)$ operators. According to \cite{Ring} 
it reads
\begin{equation}
\label{vac2}
|\tilde{0}\rangle =Ne^{\hat{Z}}|HF\rangle 
\; \; ,		
\end{equation} 	
with
\[
\hat{Z}=\frac{1}{2}\sum_{\pi ,J} \sum_{(ph),(p'h')}
Z^{\pi,J}_{(ph),(p'h')}
\left[ A^+_{(p,h)}(\pi ,J) \otimes  A^+_{(p',h')}(\pi ,J)\right]^0_0
\; \; ,
\]
and the normalization $N$ defined as
\[
N=\langle HF|\tilde{0}\rangle \;\; .
\]
This form shows clearly that the quasi-boson vacuum is a superposition of 
(2p-2h), 2 (2p-2h) ... n (2p-2h) excitations coupled to zero angular momentum 
as it should, since the total spin of the ground state is zero for the 
nuclei under consideration.  In the present work and for future applications 
to inelastic scattering we assume that the quasi-boson vacuum (\ref{vac2}) 
and the excited modes (\ref{vac1}) provide a reasonable description of the 
ground state and nuclear excitations of the target. 

At this stage it is worth pointing out that there exists another explicit 
form of the correlated ground state that has been derived \cite{Dec2} by 
summing up the RPA diagram to all orders. This important work shows that the 
resulting ground state, denoted here as $|RPA\rangle $, has exactly the same 
structure as the quasi-boson vacuum, but it reveals also that the quasi-boson 
counts twice the lowest order term of the perturbation theory. How it affects 
mean values of one body operator is now shown on the matrix elements of the 
one-body density operator.

\subsubsection{One-body density matrix for the RPA ground state}
\label{obdme}

The one-body density matrix calculated in correlated ground states is no 
longer diagonal but contains all the elements of the form $\rho_{h,h'}$ and 
$\rho_{p,p'}$. The non-diagonal particle-hole matrix elements vanish because 
of the structure of the ground state. Besides, on the account of symmetries it can be shown that the density matrix reduces to diagonal block matrices 
labeled by $(l,j,\tau)$ and independent of the projection $m$ of the angular 
momentum $j$, namely
\begin{equation}
\label{obdme1}
\rho_{(\alpha),(\beta)}=
\delta_{l_{\alpha},l_{\beta}}	\; \delta_{j_{\alpha},j_{\beta}} 
\; \delta_{\tau_{\alpha} , \tau_{\beta}} \; 
\rho_{(n_{\alpha} ,l_{\alpha} ,j_{\alpha} ,
\tau_{\alpha}), (n_{\beta} ,l_{\alpha} ,j_{\alpha} ,\tau_{\alpha})}
\; \; .
\end{equation} 
Finally, it is often convenient in the formalism to perform the summation 
over $m$ in advance and to consider the following quantities instead
\begin{equation}
\bar{\rho}_{(\alpha),(\beta)}=\sum_m \rho_{(\alpha),(\beta)}
=(2j_{\alpha}+1)\rho_{(\alpha),(\beta)}
\; \; .
\end{equation}

We next provide expressions for these quantities in the cases of the quasi-boson vacuum and RPA vacuum

\begin{equation}
\bar{\rho}_{(\alpha),(\beta)}= \langle \tilde{0}|
\sum_m a^+_{(\beta)}a_{(\alpha)}|\tilde{0}\rangle  
\;\;\; , \;\;\;
\bar{\rho}_{(\alpha),(\beta)}^{RPA}= \langle RPA|
\sum_m a^+_{(\beta)}a_{(\alpha)}|RPA\rangle 
\;\; .
\end{equation} 

The calculation in the quasi-boson vacuum is straightforward. We only give 
the result for the particle and hole cases, respectively, as

\begin{equation}
\label{rpaobdme}
\bar{\rho}_{(\alpha),(\beta)}=\delta_{(\alpha),(\beta)} 
\sum_{i,J,\pi,h} (2J+1)Y^{\pi,J}_{i,(\alpha ,h)} 
Y^{\pi,J}_{i,(\beta ,h)} \delta_{\tau_h , \tau_{\alpha}}
\;\; ,
\end{equation} 
and
\[
\bar{\rho}_{(\alpha),(\beta)}=
\delta_{(\alpha),(\beta)} \left[\delta_{n_{\alpha},n_{\beta}} 
-\sum_{i,J,\pi,h} (2J+1)Y^{\pi,J}_{i,(\alpha ,h)} 
Y^{\pi,J}_{i,(\beta ,h)} \delta_{\tau_h , \tau_{\alpha}} \right]
\;\; ,
\]
with the definition  $\delta_{(\alpha),(\beta)}=\delta_{l_{\alpha},l_{\beta}}
\delta_{j_{\alpha},j_{\beta}}\delta_{\tau_{\alpha},\tau_{\beta}}$.

In order to calculate the RPA one-body matrix elements one refers oneself to \cite{Dec1} 
where expressions of the occupation probabilities of single particle orbital 
in the RPA state can be found. Although such probabilities involve only 
diagonal matrix elements of the density, it is not difficult to generalize 
an expression for the non-diagonal ones. It turns out that the  one-body matrix elements
in the RPA state and those in the quasi-boson vacuum (\ref{rpaobdme}) differ 
only by the lowest order contribution in the perturbation theory. The 
correction terms are given for particle and hole cases, respectively, as
\begin{equation}
\label{obdmedbl}
\Delta\bar{\rho}_{(\alpha),(\beta)}=-\frac{1}{2}\delta_{(\alpha),(\beta)} \sum_{J,\pi}(2J+1)\sum_{p',h',h}\frac{B^{\pi,J}_{(\alpha,h),(p',h')}B^{\pi,J}_{(\beta,h),(p',h')}\delta_{\tau_{\alpha},\tau_h}}{(\epsilon_{(p',h')}+\epsilon_{(\alpha ,h)})(\epsilon_{(p',h')}+\epsilon_{(\beta ,h)})}
\; \; ,
\end{equation}
and
\[
\Delta\bar{\rho}_{(\alpha),(\beta)}=-\frac{1}{2}\delta_{(\alpha),(\beta)} \sum_{J,\pi}(2J+1)\sum_{p,p',h'}\frac{B^{\pi,J}_{(p,\alpha),(p',h')}B^{\pi,J}_{(p,\beta),(p',h')}\delta_{\tau_{\alpha},\tau_p}}{(\epsilon_{(p',h')}+\epsilon_{(p,\alpha)})(\epsilon_{(p',h')}+\epsilon_{(p,\beta)})}
\; \; ,
\]
where the $\epsilon_{(p,h)}=\epsilon_p-\epsilon_h$ are the free particle-hole 
pair energies, and $B^{\pi,J}_{(p,h),(p',h')}$ the values defined in 
(\ref{Bdef}) for particle-hole pairs with good angular momentum $J$ and parity 
$\pi$. With these notations, the RPA density matrix reads

\begin{equation}
\label{obdmerpa}
\bar{\rho}_{(\alpha),(\beta)}^{RPA}=\bar{\rho}_{(\alpha),(\beta)}+\Delta\bar{\rho}_{(\alpha),(\beta)}
\; \; .
\end{equation}
This expression is folded with the Melbourne g-matrix  (see (\ref{uopt}))
and the optical potential so obtained is then used to calculate elastic 
scattering observables.

\subsubsection{Structure of correlated ground states}
\begin{figure}
\begin{center}
\end{center}
{\includegraphics[scale=.45]{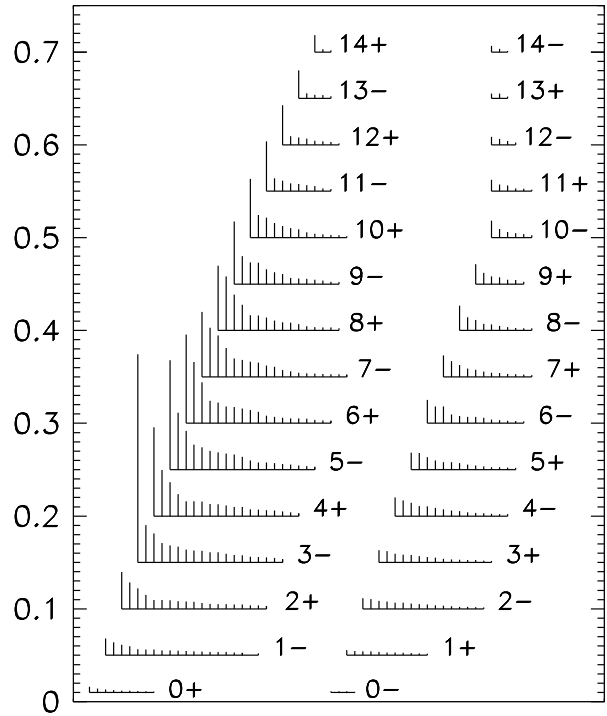}}
\caption{Values of the quantities $\bar{\theta}^{\pi,J}_{\alpha}$ defined in 
the text. We present the contributions for $^{208}$Pb, $\alpha$ = 1 $\rightarrow$ 20 first states of each ($\pi$,J) block. }	
\label{ratios}
\end{figure}

From inspection of the vacuum structure (\ref{vac2}) as outlined in Appendix \ref{appen2}, it is clear that the 
$\theta_{\alpha}$  amplitudes (see \ref{vacstruc}) provide a 
direct measure of ground state correlations. Taking into account the 
$(2J+1)$-fold degeneracy of the $\theta_{\alpha}$ 's in each $(\pi ,J)$ 
subspace, the ratio
\begin{equation} 
\bar{\theta}^{\pi ,J}_{\alpha}=\frac{(2J+1)}{(2J_{Ref}+1)} \theta^{\pi,J}_{\alpha}
\; \; ,
\end{equation} 
is a measure of the relative importance of each subspace, with J$_{Ref}$ taken
 as the multipolarity of the one which provides the main contribution to the 
overall correlations (here, J$_{Ref}$ = 3). These ratios shown in 
Fig.\ref{ratios} for $^{208}$Pb indicate that some natural and unnatural 
parity states of all $(\pi ,J)$ subspaces, even high spin ones, are worthy of 
consideration for building the correlated ground state.

\begin{figure}
\begin{center}
{\includegraphics[scale=.5]{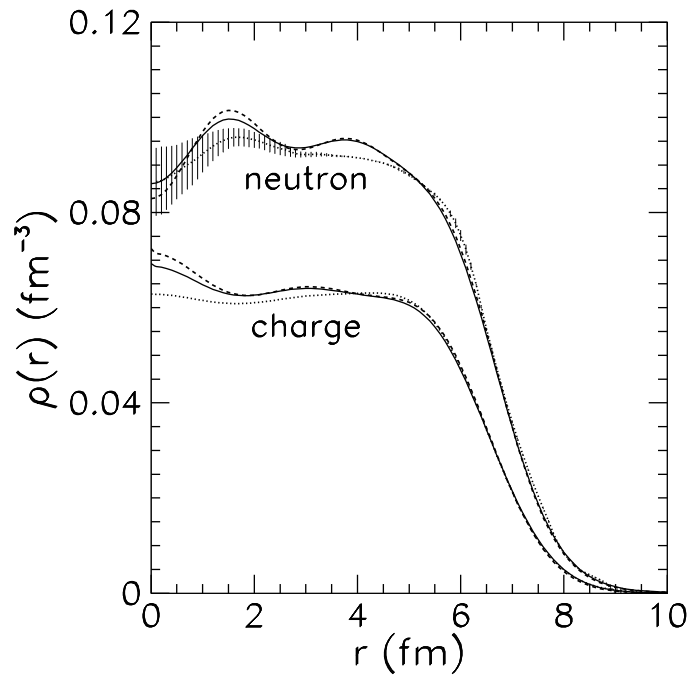}}
\end{center}
\caption{Charge and neutron radial densities of $^{208}$Pb. Comparisons
between experimental data \cite{Froi,Star} (dotted curves), correlated 
(full curves) and uncorrelated (dashed curves) calculations.}	
\label{raddens}
\end{figure}

As the correlations are smearing out the occupation probability distribution 
of proton and neutron single-particle levels around their respective Fermi 
energies, the radial g.s. densities get depleted towards the nuclear center. 
This effect can be seen in Fig.\ref{raddens} where measured charge and neutron
 distributions are shown together with our HF and HF+RPA predictions for 
$^{208}$Pb. Calculated root mean square (rms) radii of proton, charge and 
neutron distributions as well as neutrons skins are gathered in Table 
\ref{rdata} for $^{208}$Pb as well as for $^{16}$O, $^{40}$Ca and 
$^{48}$Ca. A good overall agreement between the RPA predictions and 
experimental values is obtained.

\begin{table}
\begin{center}
\begin{tabular}{cccccccccc} \hline\hline

 Nucleus & &  $<r^2_p>^{1/2}  $  &  & $<r^2_{ch}>^{1/2}  $ &&  &  $<r^2_n>^{1/2}$  &  &  $\Delta r_{np}$  \\
& & (fm) && (fm) &&& (fm) && (fm) \\ \hline
& exp & &  & 2.730(25) \cite{Vrie}&  & & &  \\  
 $^{16}$O & HF & 2.669   & & 2.718  && & 2.647  & & -0.022 \\
& HF+RPA & 2.658  & & 2.728  && & 2.678  & & -0.020 \\  \hline
& exp &  & & 3.482(25) \cite{Vrie} && &3.312(2) \cite{Clark}&  & 
\begin{tabular}{c}
-0.040 \cite{Trzc}\\ 
-0.065(2) \cite{Clark}
\end{tabular} \\  
 $^{40}$Ca & HF & 3.408   & & 3.470  && & 3.365  & & -0.043 \\
& HF+RPA & 3.421  & & 3.483  && & 3.381  & & -0.040 \\  \hline
& exp &  & & 3.470(9) \cite{Vrie} && & 3.436(23) \cite{Clark}& & 
\begin{tabular}{c}
+0.128 \cite{Trzc}\\
+0.079(23) \cite{Clark}
\end{tabular} \\  
 $^{48}$Ca & HF & 3.441   & & 3.496  && & 3.588  & & +0.144 \\
& HF+RPA & 3.455  & & 3.510  && & 3.590  & & +0.130 \\  \hline
& exp &  & & 5.503(7) \cite{Vrie} && & 5.511(11) \cite{Clark} & & 
\begin{tabular}{c}
+0.15(2) \cite{Trzc}\\
  +0.12(7) \cite{Csat}\\
+0.097(14) \cite{Clark}
\end{tabular} \\
 $^{208}$Pb & HF & 5.432   & & 5.475  && & 5.567  & & +0.135 \\
& HF+RPA & 5.467  & & 5.504  && & 5.592  & & +0.125 \\  \hline\hline
 \end{tabular}
\end{center}
\caption{ Proton, charge, and neutron rms radii for $^{16}$O, $^{40}$Ca, $^{48}$Ca and $^{208}$Pb.
Comparisons between present HF and HF+RPA predictions, and experimental 
values . The neutron skin $\Delta r_{np}$ is defined as 
$\Delta r_{np}=\langle r^2_n \rangle ^{1/2}- \langle r^2_p \rangle ^{1/2}$. 
The estimated  $\langle r^2_n \rangle ^{1/2}$ and $\Delta r_{np}$ values of 
Refs. \cite{Clark,Trzc} are from systematics.}
\label{rdata}
\end{table}

\section{Analyses of scattering observables}
\label{analyse}
In order to test the predictions of the OMP described above, an incident proton experimental 
database  was built, comprising differential cross sections 
$\sigma (\theta ) / \sigma_{Ruth}$, analyzing powers $A_y(\theta )$ and 
spin rotation functions $R(\theta )$ and $Q(\theta )$.  References to 
these data are provided in Table \ref{database} only for $^{208}$Pb. The 
incident energies of present interest are limited to the 40-201 MeV range 
where the Melbourne OMP is most successful \cite{Am00}. For all the 
comparisons between model predictions and experimental data shown below the 
continuous and dashed curves represent the OMP calculations based on one-body 
density matrix elements of correlated (RPA) and uncorrelated ground states 
(HF), respectively. 
\begin{table}
\begin{center}
\begin{tabular}{cccccc} \hline\hline
  Energy (MeV) & Ref. & Energy (MeV) & Ref. \\ \hline

  40 & \cite{Blum} & 104.4,121.2 & \cite{Nada} \\
      45,47.3 & \cite{Oers} & 156  & \cite{Comp} \\
      61.4 & \cite{Fulm} & 160 & \cite{Ross} \\
       65  & \cite{Saka} & 182.4 & \cite{Nada} \\ 
      79.9 & \cite{Nada} & 185 & \cite{Schw} \\ 
      97   & \cite{Schw} & 201 & \cite{Hutc,Ju,Otte,Lee}\\ 
\hline\hline
 \end{tabular}
\end{center}
\caption{ $\sigma (\theta )/ \sigma _{Ruth}$, $A_y(\theta )$, $R(\theta )$ and $Q(\theta )$ database for proton scattering off $^{208}$Pb.}
\label{database}
\end{table}

\subsection{Incident protons}
\label{dataanalyse}
Proton scattering experiments have provided a wealth of valuable information 
on angular distributions for various observables at many incident energies.
 For this reason, the proton database we have formed in Table \ref{database} 
serves as the main play ground for detailed OMP analyses. We also show some 
illustrations for the three other stable doubly-closed-shell nuclei 
$^{16}$O, $^{40}$Ca and $^{48}$Ca. 

\subsubsection{$^{208}$Pb}
\begin{figure}[t]
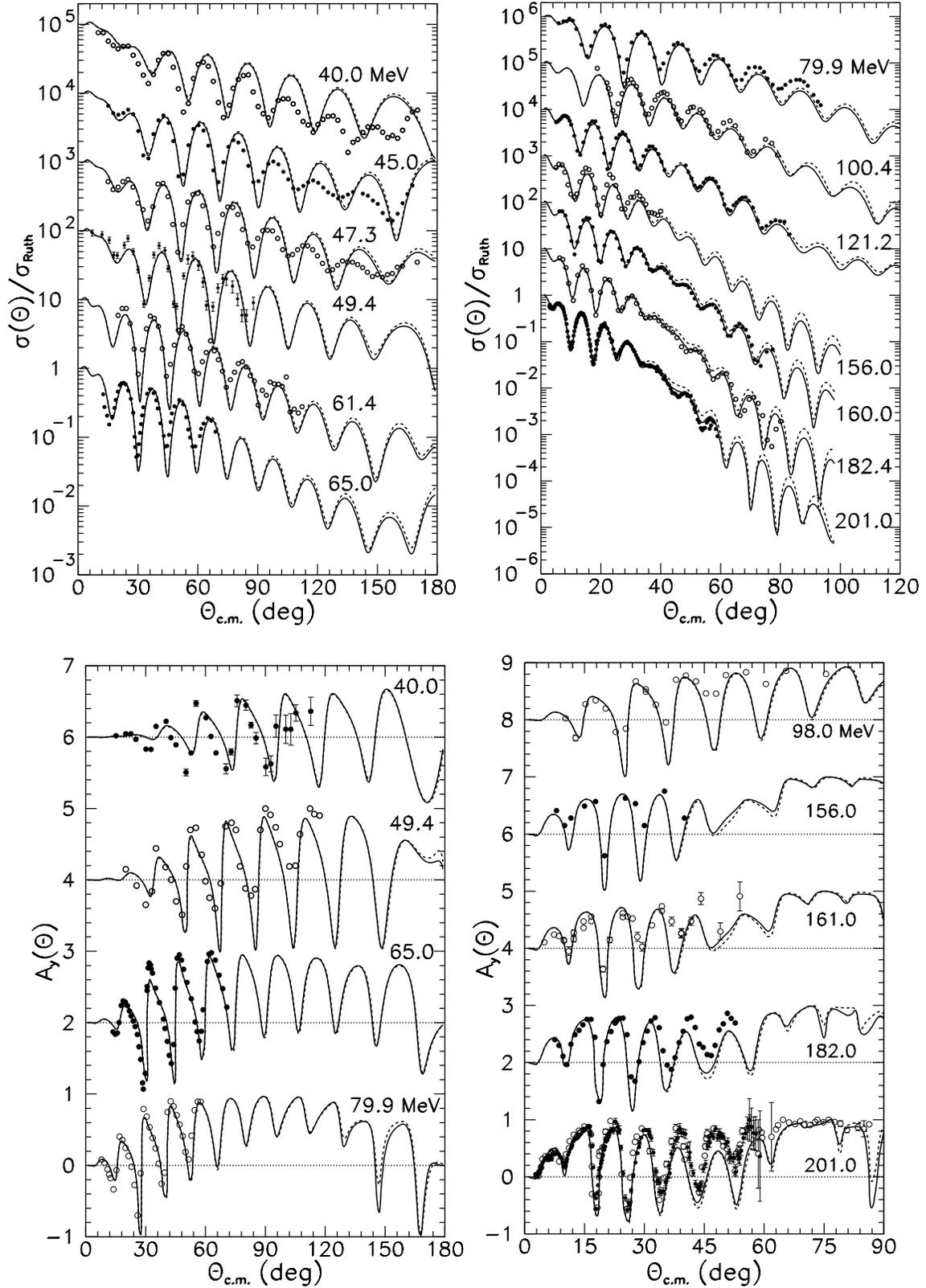

\begin{center}
{\includegraphics[scale=.45]{3a.epsi}}
{\includegraphics[scale=.45]{3b.epsi}}
\end{center}
\begin{center}
{\includegraphics[scale=.45]{3c.epsi}}
{\includegraphics[scale=.45]{3d.epsi}}
\end{center}
\caption{Differential cross sections $\sigma (\theta )/ \sigma _{Ruth}$ 
and analyzing powers $A_y( \theta )$ for protons incident on $^{208}$Pb. 
Comparison between data (symbols) and OMP predictions based on correlated 
(solid curves) and uncorrelated (dashed curves) descriptions of ground 
state. Cross sections are offset by factors 10, while analyzing powers are 
shifted by 2.}	
\label{seeppb}
\end{figure}

 The differential cross sections to be discussed below are normalized to 
Rutherford scattering cross sections to magnify differences existing between 
our OMP predictions and scattering data. This comparison is shown in the 
upper panel of Fig.\ref{seeppb}. Similar comparisons for $A_y(\theta )$ are 
shown in the lower panel of Fig.\ref{seeppb}.

For the comparison between solid (RPA-based) and dashed (HF-based) curves 
for cross sections, it turns out that the former is systematically lower 
over most scattering angles. Compared to OMP predictions based on the 
Hartree-Fock ground state density matrix, those using the RPA one are all 
in closer agreement with the spread of cross section data except maybe at 
lower incident energies where the calculated minima seem too deep. This is 
a known low energy shortcoming of the g-folding model which has been 
discussed previously \cite{Deb2000}. Nevertheless, the agreement between 
RPA-based calculations and measured differential cross-sections is 
spectacular, especially at the higher incident energies, where HF- and 
RPA-based calculations differ the most.
Extending the comparison from experimental cross sections to analyzing 
powers, it may be seen in the lower part of Fig.\ref{seeppb} that the 
correlated ground state specifications lead to a excellent overall OMP 
description of the $A_y(\theta )$ data spread, especially at medium 
angles for energies E $\ge$ 150 MeV.

\begin{figure}[t]
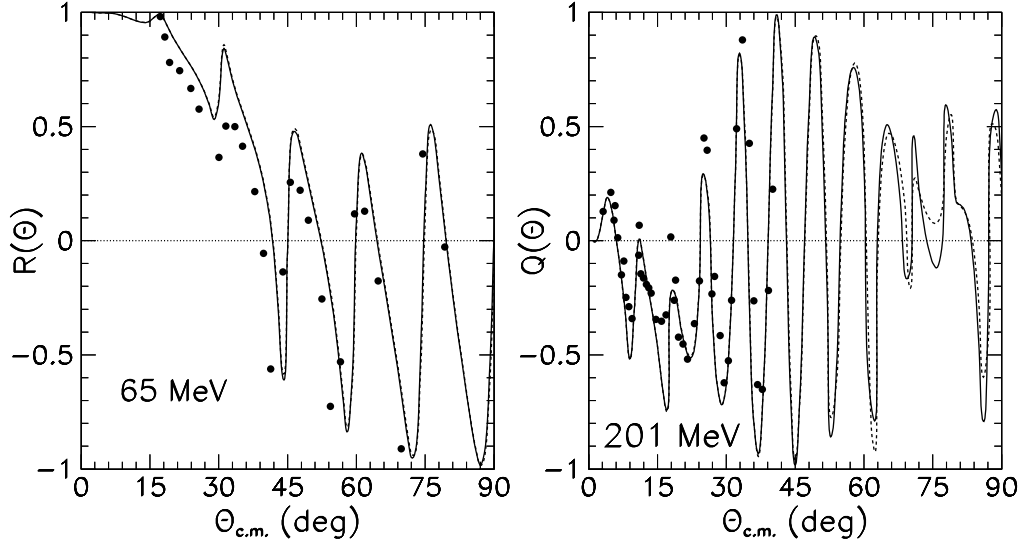

\begin{center}
{\includegraphics[scale=.4]{4a.epsi}}
{\includegraphics[scale=.4]{4b.epsi}}
\end{center}
\caption{Spin rotation functions R($\theta$) and Q($\theta$) at 65 
and 201 MeV for protons incident on $^{208}$Pb. Comparisons between 
experimental data (symbols) and present OMP calculations for correlated 
(solid curves) and uncorrelated (dashed curves) ground state descriptions.}
\label{qandr}
\end{figure}

A similar statement is made for the spin-rotation functions $R(\theta )$ 
and $Q(\theta )$ measured at 65 and 201 MeV, respectively. As can be seen 
in Fig.\ref{qandr} the phasing and amplitude of these measured observables 
are well accounted for by our OMP calculations, although  these observable 
predictions do not seem very sensitive to RPA correlations.


\subsubsection{Other doubly-magic nuclei}
\begin{figure}[t]
\begin{center}
{\includegraphics[scale=.4]{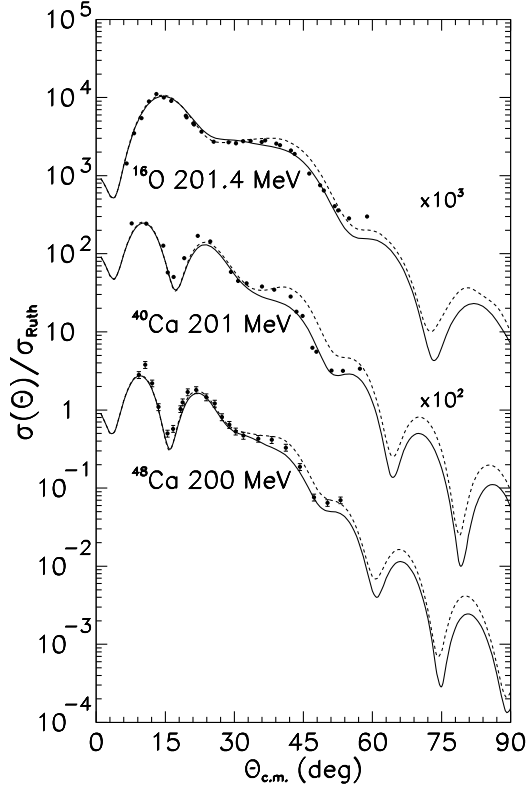}}
\end{center}
\caption{Differential cross sections $\sigma (\theta )/ \sigma _{Ruth}$  
for protons incident on $^{16}$O, $^{40}$Ca and $^{48}$Ca. Comparison 
between experimental data (symbols) and OMP predictions based on correlated 
(solid curves) and uncorrelated (dashed curves) descriptions of ground 
state. Cross sections offset factors and proton incident energies are 
indicated on the figure. Data are taken from Ref. \cite{Seif} for $^{16}$O 
and $^{40}$Ca and from  \cite{Feldm} for$^{48}$Ca.}
\label{figOCa}
\end{figure}

Calculations were also performed for protons incident on the other stable 
doubly-magic nuclei $^{16}$O, $^{40}$Ca and $^{48}$Ca. Although these 
calculations were performed for all incident energies where experimental 
data are available, Fig.\ref{figOCa} only displays comparisons at highest 
energies, where the difference between HF- and HF+RPA- based OMPs is the 
most striking. Those results are representative of the agreement obtained 
over the 60-201 MeV range. For these doubly magic nuclei, comparison between 
calculations using correlated and uncorrelated ground state density 
matrices, and the experimental data, allows us to confirm the conclusions 
of the $\vec{\mbox{p}}$+$^{208}$Pb scattering study made above with a larger 
data sample.


\subsection{Incident neutrons}
Although some neutron scattering data is available \cite{Osbo,Klug,Klug2} 
at neutrons energies higher than 40 MeV, those data sets (with the notable 
exceptions of \cite{Klug,Klug2}) do not extend far enough in angles to allow 
for discrimination between the nuclear structure models used as a basis 
for our OMP analyses. Thus, those datasets can be described in a satisfactory 
way by our OMP using either HF or RPA one-body density matrix. Moreover, 
when  comparing incident proton and incident neutron calculations, no effect 
specific to incident neutrons was observed,  and like for incident protons, 
the RPA-based neutron-nucleus OMP calculations predict cross sections that 
are systematically lower at large angles than their HF counterparts. 
Nevertheless, the scarcity of high energy, large angular range neutron
 scattering data, calls for new  measurements of the quality of those 
in \cite{Klug,Klug2}, maybe at higher energy. 

We conclude this analyses with making the statement that the RPA correlations 
have sizeable impacts on the OMP predictions only at the higher incident 
energies of present interest and for center-of-mass scattering angles larger 
than typically $\theta \sim$30$^o$. This statement is relevant to 
$^{16}$O,  $^{40}$Ca, $^{48}$Ca and $^{208}$Pb target nuclei.
\subsection{Discussion}

\subsubsection{Probing ground state correlations}
\begin{figure}[t]
\begin{center}
{\includegraphics[scale=.45]{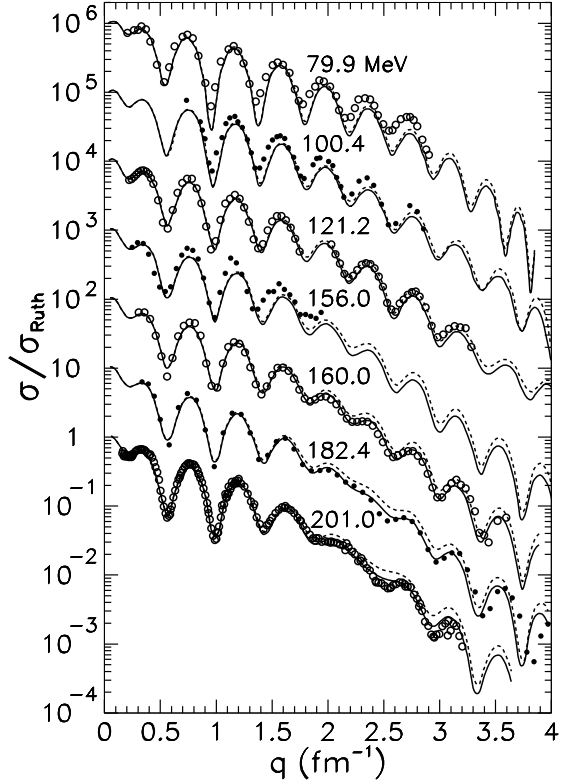}}
\end{center}
\caption{Proton elastic scattering from $^{208}$Pb: differential cross 
sections $\sigma / \sigma _{Ruth}$ as functions of the momentum 
transfer $q$. For more details, see caption of Fig. \ref{seeppb}} 	
\label{seeq}
\end{figure}

In Sec. \ref{dataanalyse}, we have shown that similarly to electron 
scattering, nucleon scattering is sensitive to small details of the 
nuclear structure of the target nuclei, such as those stemming from the 
presence of long range correlations in the target ground state. Moreover, 
including such correlations do improve the agreement 
between calculated and measured scattering cross sections. Next comes 
the difficult question of identifying the features of the correlated 
density matrix that nucleon scattering is sensitive to.  
Looking at  Fig.\ref{seeppb} can provide us with hints to that effect: 
the differences between HF- and RPA-based calculations can be seen 
to be stronger at large angles, suggesting that such differences appear 
when more interior regions of the target are probed.   Re-plotting the 
p+$^{208}$Pb scattering cross sections as functions of the momentum 
transfer $q$ (see Fig.\ref{seeq}) confirms that indeed, for all energies, 
differences between HF- and RPA-based calculations are associated with 
values of $q$ larger than 1.7 fm$^{-1}$, and thus deeper regions of the 
target. Fig.\ref{raddens} displays the radial charge density of $^{208}$Pb 
calculated with (solid curve) and without (dashed curve) RPA correlations 
in the ground state, showing the well known effect of RPA correlations, 
i.e. depleting the interior of the density distributions and enlarging 
the distributions rms radii. The fact that only $q \geq 1.7$ fm$^{-1}$  
cross sections are affected by RPA correlations suggests that  this value 
of the momentum transfer constitutes the threshold above which the depletion 
of the inner regions of the target becomes sizable. However, since the 
density matrix used as an input to our microscopic OMP calculations conveys 
much more complex  nuclear structure information that the radial density 
alone, disentangling  the effects of the RPA correlations on nucleon 
scattering is a much more difficult task than the   analysis of  the $q$ 
dependence of the radial density. Therefore, unlike the case of  electron 
scattering,  such an analysis can at best  provide  qualitative insight 
into the actual sensitivity of nucleon scattering to the presence of RPA 
correlations in the one-body density matrix of the target.

\subsubsection{Double counting}
Further tests of the sensitivity of our scattering predictions to changes 
in  matter distributions have been performed by ignoring the  $\Delta \rho$  
double counting correction terms (see (\ref{obdmedbl}) and (\ref{obdmerpa})). 
Elastic scattering calculation results performed with (solid curve) and 
without (dashed curve) these correction terms
are shown in Fig.\ref{pb201} for 201 MeV protons incident on $^{208}$Pb. 
First, Fig.\ref{pb201} shows that including or ignoring the   $\Delta \rho$ 
correction produces non-negligible changes in the calculated scattering cross 
section. Moreover, except for a local improvement at $\theta =$ 54$^o$ over 
those  using $\Delta \rho \neq 0$  (solid curve), the agreement between data 
and the OMP calculation with $\Delta \rho = 0$ is worse all over the range 
$\theta \geq$ 34$^o$. Setting $\Delta \rho$ to 0 leads to increasing the rms 
radii from $\langle r_{ch}^2\rangle ^{1/2}$ = 5.504 fm ($\Delta \rho \neq 0$, 
see Table \ref{rdata}) to $\langle r_{ch}^2\rangle ^{1/2}$ = 5.517 fm 
($\Delta \rho = 0$), a value falling apart from the experimental result 
$\langle r_{ch}^2\rangle ^{1/2}$ = 5.503(7) fm (see Table \ref{rdata}). 
The $^{208}$Pb neutron and proton radial shapes calculated assuming 
$\Delta \rho = 0$ (dotted curves) and $\Delta \rho \neq 0$ (full curves) 
are shown in the insert of Fig.\ref{pb201}. 
The above discussion shows that the $\Delta \rho$  double counting correction 
to the RPA density matrix should not be ignored in scattering calculations.

\subsubsection{Skyrme Hartree-Fock model}
\begin{figure}
\begin{center}
{\includegraphics[scale=.6]{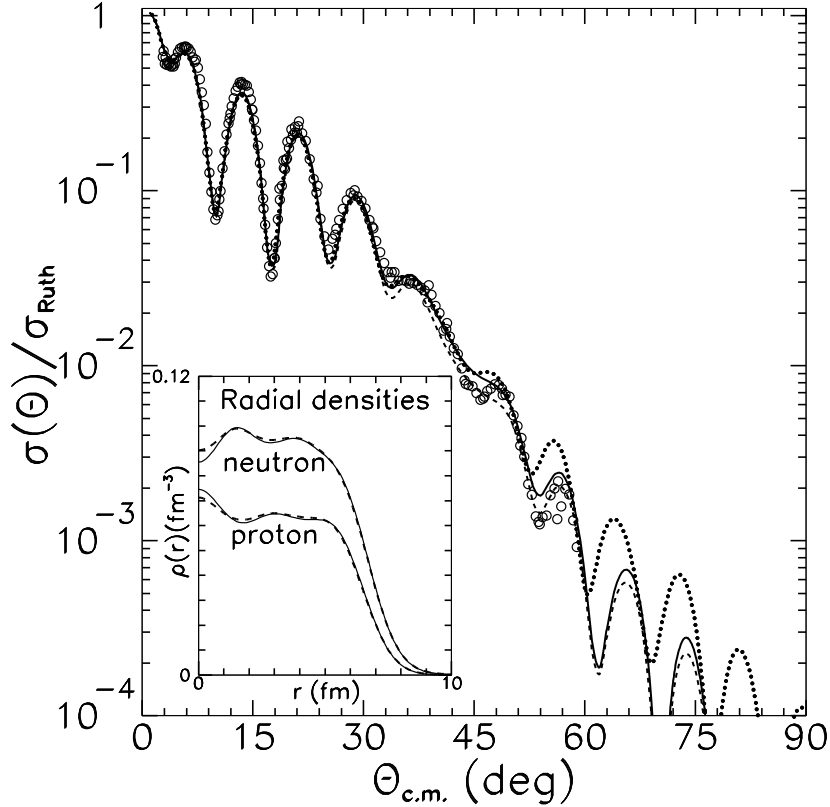}}
\end{center}
\caption{Differential cross sections $\sigma (\theta )/ \sigma _{Ruth}$  
for 201 MeV protons incident on $^{208}$Pb. Comparison between experimental 
data (symbols) and OMP predictions based on correlated (solid curves), 
correlated without double counting corrections (dashed curves), and 
Hartree-Fock SkM* (dotted curve) descriptions of ground state. The insert 
shows comparison between proton and neutron radial densities for correlated 
(solid curves) and correlated without double counting corrections 
(dashed curves) descriptions of ground state.}
\label{pb201}
\end{figure}

In recent years, Skyrme Hartree-Fock models have been considered to assess 
the neutron rms radius in $^{208}$Pb \cite{Ka02}. Furthermore, various Skyrme 
force parameterizations have been tested in NA g-folding model calculations 
to discern which one provides the best representation of the neutron density. 
As a result, it turns out that SkM$^*$ seems appropriate when combining 
analyses of electron and nucleon scattering data. g-folding model 
calculations with HF/SkM$^*$ as input have again been performed and compared 
with calculations based on the present correlated ground state densities. 
The comparison made for (p,p) scattering off $^{208}$Pb at 201 MeV is shown 
in Fig.\ref{pb201} where the dotted and solid curves are for results from 
the HF/SkM$^*$ and HF+RPA/D1S based OMPs, respectively. The dotted and solid 
curve overlap each other over most of the angular range, except perhaps for 
angles above 50$^o$. This is not surprising since both HF/SkM$^*$ and 
HF+RPA/D1S structure calculations provide nearly identical radial matter 
distributions and neutron skins for $^{208}$Pb.  However this similarity 
conceals more fundamental differences: whereas the SkM$^*$ interaction was 
designed to reproduce the measured charge radii of many stable nuclei within 
the HF framework only (its parameters take care of correlations present in 
nuclear ground states in an effective way at the mean field level), the D1S 
interaction is designed not to include such correlation effects in its 
parameterization, so that correlations can be explicitly taken care of, in a 
detailed  way, at a level that goes beyond that of the mean field 
approximation. 

\section{Conclusions}
We present a comprehensive analysis of ground state structure properties 
of doubly-closed-shell nuclei, together with the impacts they have on the 
interpretation of nucleon elastic scattering observables within the 
Melbourne g-folding model. Long range correlations are treated in the 
self-consistent RPA theory implemented with the D1S force, and the 
longstanding problem relevant to double counting is solved to calculate 
local and non-local densities. The theoretical framework which in the past 
proved successful in the interpretation of electron scattering measurements 
is shown to be equally successful in the analyses of nucleon elastic 
scattering up to 201 MeV. All the measured differential cross-sections, 
analyzing powers and spin-rotation functions are well described, 
{\em without any adjusted parameter}. Turning off RPA correlations 
(or not implementing them properly, i.e. without considering double counting corrections) 
negatively affects the agreement between experimental data and calculations, 
an effect which gets more and more sizeable as incident energy and momentum 
transfer increase.  It seems plausible that the differences observed between 
predictions are strongly tied to differences between correlated and 
uncorrelated matter densities at the surface and also towards nuclear center. 
Finally, since in the RPA theory, the correlated ground state happens 
to be the vacuum on which excited states are built as quasi-bosons 
excitations,  a framework is at hand for extending our g-folding model 
analyses from elastic scattering to inelastic scattering from low to high 
excitation energy levels. Work along this line is in progress.
\label{finish}

\begin{acknowledgments}
We would like to acknowledge the usefulness of discussions with
S. Peru on the RPA formalism and codes. We are also
deeply indebted to J. Raynal for his  relentless support of his
microscopic DWBA code and for invaluable insights into many
obscure but nevertheless very important points.
\end{acknowledgments}

\appendix
\section{Definition.}
\label{appen1}
The Hartree-Fock solutions in the spherical case take the form
\begin{equation}
\langle x|(nlj),m,\tau\rangle =R^{\tau}_{nl}(r) i^l \left[\chi^{1/2}(\sigma)\otimes Y^l(\Omega) \right]^j_m\chi^{1/2}(\tau)
\; \; .
\end{equation}
The operator $a^+_{(nlj),m,\tau}$ creates a particle in this state and 
its hermitian conjugate defines the destruction operator $a_{(nlj),m,\tau}$. 
It is convenient to define destruction operators $\bar{a}_{(nlj),m,\tau}$ 
through the relation
\begin{equation}
\bar{a}_{(nlj),m,\tau}=(-)^{j+m}a_{(nlj),-m,\tau}
\; \; .
\end{equation}

Indeed, with this definition, both $a^+_{(nlj),m,\tau}$  and 
$\bar{a}_{(nlj),m,\tau}$  transform under rotations like the component m of 
an irreducible tensor of rank j . Consequently creation operators of 
particle-hole pairs of definite angular momentum are readily constructed 
with the usual rules for coupling two tensors:
\begin{equation}
A^+_{(p,h)}(\pi,J,M)=\left[ a^+_{(p),\tau} \otimes \bar{a}_{(h),\tau} \right]^J_M=\sum_{m_p,m_h}C^{j_p \;\;\; j \;\;\; J}_{m_p  m_h  M}
 \; a^+_{(p),m_p,\tau} \; \bar{a}_{(h),m_h,\tau}
\; \; .
\end{equation}
The parity ``$\pi$'' of the particle-hole pair that we indicate explicitly 
is defined by: $\pi=(-)^{l_p-l_h}$. As we did for the fermions, we define 
operators $\bar{A}$
\begin{equation}
\bar{A}_{(p,h)}(\pi,J,M)=(-)^{J-M}A_{(p,h)}(\pi,J,-M)
\; \; ,
\end{equation} 
which annihilate particle-hole pairs of angular momentum J and projection M. 
As a consequence, by mixing $A^+$ and $\bar{A}$, the Bogolyubov 
transformation defines operators $\Theta^+$ and $\Theta$  which respectively 
create and annihilate collective modes of definite angular momentum and 
parity. Finally, let us also recall that we consider only neutron and 
proton particle-hole pairs and consequently $\tau_p=\tau_h$.

\section{The quasi-boson vacuum.}
\label{appen2}
The expression of the quasi-boson vacuum takes a very simple form in the 
so-called canonical representation defined as follows
\begin{equation}
B^+_{\alpha}(\pi,J,M)=\sum_{(ph)}D^{\pi,J}_{\alpha,(ph)}A^+_{(ph)}(\pi,J,M)
 \; \; ,
\end{equation} 
\[
\bar{B}_{\alpha}(\pi,J,M)=\sum_{(ph)}D^{\pi,J}_{\alpha,(ph)}\bar{A}_{(ph)}(\pi,J,M)
\; \; .
\]
The transformation D is an orthogonal transformation which mixes separately 
the creation and destruction operators of the original particle-hole pairs 
(it is orthogonal because our Bogolyubov transformation is real). It is 
defined by solving the eigen-values problem
\begin{equation}
\sum_{(p'h')}\left[\tilde{Y}^{\pi,J}Y^{\pi,J}\right]_{(ph),(p'h')}D^{\pi,J}_{\alpha,(p'h')}=\rho^{\pi,J}_{\alpha}D^{\pi,J}_{\alpha,(ph)}
\; \; ,
\end{equation} 
\[
\left[\tilde{Y}^{\pi,J}Y^{\pi,J}\right]_{(ph),(p'h')}=\sum_i Y^{\pi,J}_{(ph)}Y^{\pi,J}_{(p'h')}
\; \; .
\]
In this representation, the vacuum reads
\[
|\tilde{0}\rangle =\prod_{\pi,J} \left( \prod_{\alpha} \mbox{ch}\theta^{\pi,J}_{\alpha} \right)^{(2J+1)}e^{\hat{Z}}|HF\rangle 
\; \; ,
\]
with
\begin{equation}
\label{vacstruc}
\hat{Z}=\frac{1}{2}\sum_{\pi,J,\alpha} \mbox{th} \theta^{\pi,J}_{\alpha} \sum_M B^+_{\alpha}(\pi,J,M) \bar{B}_{\alpha}(\pi,J,M)
=\frac{1}{2}\sum_{\pi,J,\alpha} \mbox{th} \theta^{\pi,J}_{\alpha} \hat{J}\left[B^+_{\alpha}(\pi,J)\otimes \bar{B}_{\alpha}(\pi,J)\right]^0_0
\; \; .
\end{equation}

The angle $\theta^{\pi,J}_{\alpha}$ is related to the eigenvalues 
$\rho^{\pi,J}_{\alpha}$ through the relation:
\begin{equation}
\mbox{th} \theta^{\pi,J}_{\alpha}=\sqrt{\frac{\rho^{\pi,J}_{\alpha}}{1+\rho^{\pi,J}_{\alpha}}}
\; \; .
\end{equation}
This form shows clearly that the $\theta^{\pi,J}_{\alpha}$ 's provide 
a direct measure of the correlations which are induced by the RPA modes.

\bibliography{dmix}


\end{document}